# Atom probe analysis of BaTiO$_3$ enabled by metallic shielding

Se-Ho Kim[a,b,†,*], Kihyun Shin[c,d,e,†], Xuyang Zhou[a], Chanwon Jung[a], Hyun You Kim[d], Stella Pedrazzini[f], Michele Conroy[f], Graeme Henkelman[c], Baptiste Gault[a,f,*]

a Max-Planck-Institut für Eisenforschung, Max-Planck-Straße 1, 40237 Düsseldorf, Germany

b Department of Materials Science and Engineering, Korea University, Seoul 02841, Republic of Korea

c Department of Chemistry and the Oden Institute of Computational Engineering and Sciences, The University of Texas at Austin, Austin, Texas 78712, United States

d Department of Materials Science and Engineering, Chungnam National University, Daejeon 34134, Republic of Korea

e Department of Materials Science and Engineering, Hanbat National University, Daejeon, 34158, Republic of Korea

f Department of Materials, Royal School of Mines, Imperial College London, SW7 2AZ London, UK

[†] Contributed equally on this work

[*] Co-corresponding authors: s.kim@mpie.de, b.gault@mpie.de

**Abstract**

Atom probe tomography has been raising in prominence as a microscopy and microanalysis technique to gain sub-nanoscale information from technologically-relevant materials. However, the analysis of some functional ceramics, particularly perovskites, has remained challenging with extremely low yield and success rate. This seems particularly problematic for materials with high piezoelectric activity, which may be difficult to express at the low temperatures necessary for satisfactory atom probe analysis. Here, we demonstrate the analysis of commercial BaTiO$_3$ particles embedded in a metallic matrix. Density-functional theory shows that a metallic coating prevents charge penetration of the electrostatic field, and thereby suppresses the associated volume associated change linked to the piezoelectric effect.

Since the discovery of the first ferroelectric material in the 1950s, BaTiO$_3$[1] has been widely used across a range of applications for its piezoelectric properties[2] and in stacked ceramic capacitors, due to its high dielectric constant and low dielectric loss. For instance, the development of multilayer ceramic chip capacitors (MLCC) has improved volumetric efficiency, cost reduction, and performance such that their sales figures are the highest among fine-ceramic products.

Typically to fabricate a conventional MLCC, uniform-sized BaTiO$_3$ powders are cast on a conductive foil (*e.g.* Ni or Cu) and sintered. The key challenges to acquire reliable MLCCs from such a manufacturing process are having a constant dielectric layer thickness (several μm level in thickness) and the most careful control over the presence of impurities and/or dopants. Nowadays, the thickness of the capacitor cell can go down to a micron with high yield. However, the latter challenge is still a critical issue. The intrinsic properties of BaTiO$_3$ are easily modified by chemical doping. For example, less than 1 at.% of elemental modifier in the system may alter the Curie temperature and dielectric constant[3–6]. Extensive work has been reported on enhancing dielectric and ferroelectric properties of BaTiO$_3$ through Ba-site substitution using alkaline-earth[7–9] ($Mg^{2+}$, $Ca^{2+}$, $Sr^{2+}$) and rare-earth elements[10–12] ($La^{3+}$, $Ce^{3+}$, $Dy^{3+}$).

While there have been advances in analyzing constituent elements via inductively coupled plasma-optical emission spectrometry[13] and mass spectrometry[14], electron energy loss spectroscopy[15–17] and energy-dispersive X-ray spectroscopy (EDS)[18,19], time-of-flight secondary ion mass spectrometry[20], nuclear magnetic resonance spectroscopy[21], the direct, quantitative 3D elemental mapping of BaTiO$_3$ materials remains challenging. Atom probe tomography (APT) is expected to exhibit the required spatial resolution and elemental detection sensitivity. The physical principle on which the technique is based on field evaporation: an intense standing electric field ($10^{10}$ V·m$^{-1}$) facilitates the removal and ionization of surface atoms from a needle-shaped specimens; the resulting ions are then accelerated by electrostatic forces toward a position-sensitive detector that measures their time-of-flight and impact coordinates, which are used for calculating a mass spectrum and reconstructing a 3D atom map, respectively[22].

Despite its potential to understand microstructural and chemical properties in perovskites, there have been only very few reports of APT analyses. In our preliminary work on a bulk $BaTiO_3$ perovskite material, five specimens were tested but all of them failed to acquire meaningful data (see Figure S1). Inspired by the use of a metallic coating technique to prevent early specimen fracture or delithiation[23–25], here we propose a possible approach to facilitate APT measurement of piezoelectric $BaTiO_3$.

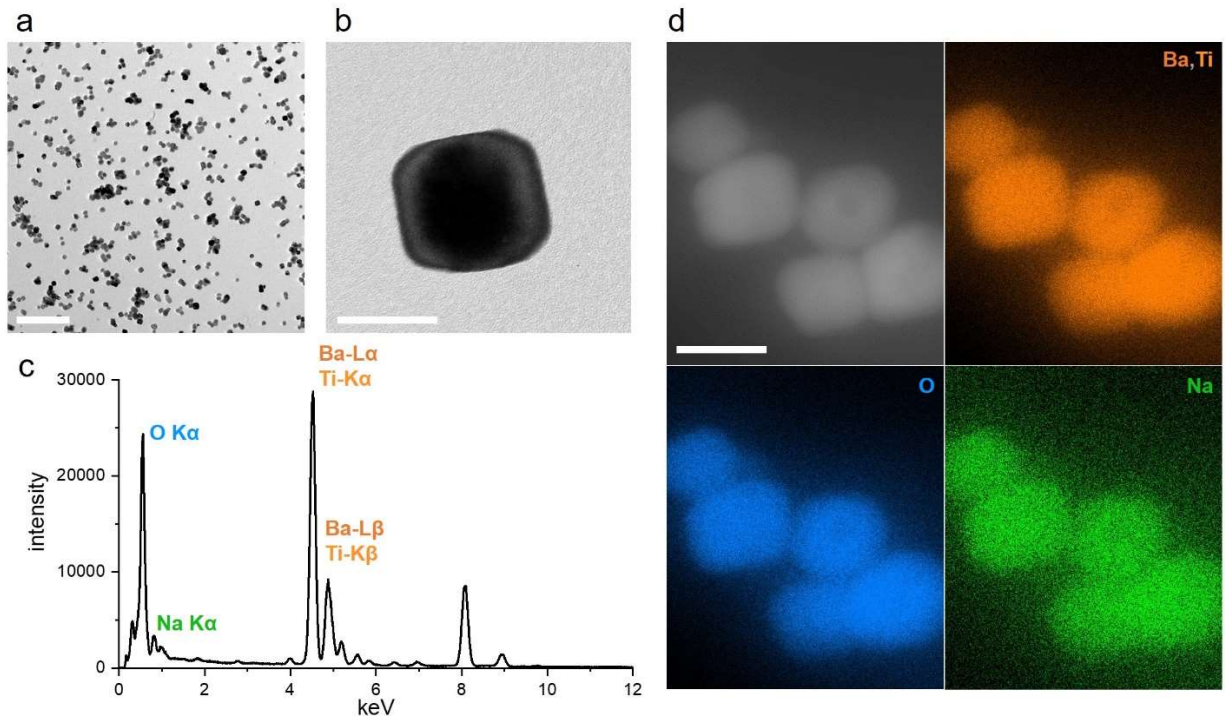

**Figure 1.** Bright field TEM images of $BaTiO_3$ nanoparticles at (a) high and (b) low magnification. (c) Acquired EDS spectrum from the particle. A significant peak intensity at 8 keV corresponds to Cu from the commerical TEM grid. (d) Maps of TEM with EDS chemical composition on the particles: Ba and Ti (orange), O (blue), and Na (green). The scale bars are (a) 500 and (b-d) 30 nm.

BaTiO$_3$ commercial powder (>99%, Sigma Aldrich) was sourced, with a specification sheet indicating un-specified impurity content of 10,000 ppm. Particles were first dispersed onto a Cu grid and imaged with a JEM-2200FS transmission electron microscopy (TEM; JEOL) operating at 300 kV equipped with an energy-dispersive X-ray analysis (EDS) system. Figures 1a and 1b show bright-field TEM images of the commercial BaTiO$_3$ particles. The particles are 30 nm in size and have a chamfered cuboidal shape. The morphology of the particles is uniform, which is important to facilitate sintering, as the compactness decreases with a broader particle size/shape distribution[26]. EDS analysis was performed (Figure 1c) with the characteristic X-ray peaks of the acquired spectrum were interpreted according to Ref.[27]. The characteristic X-ray peaks of the main constituents (Ba, Ti) overlap substantially, making distinction challenging. Yet, despite a high background, signals from Na and Sr are detected (Figure S2). Both elements are well reported to substitute Ba cation (A site in ABO$_3$ perovskite structure)[28,29]. Elemental mapping in Figure 1d show a homogenous distribution of all elements. Nevertheless, as (S)TEM is a through-thickness projection microscopy, precise localization of each element is not readily possible without deploying highly advanced, electron tomography techniques.

To facilitate APT specimen preparation, we followed the protocol described by [30] to embed the powder, Figure 2a, in a Ni-metallic matrix using a commercial nanoparticle depositor (Oxford Atomic. Ltd.), with a current of -19 mA. Ni is selected for the coating material simply because, in the MLCC manufacturing process, Ni is not only used as a capacitor electrode but also for encapsulating materials for sintered BaTiO$_3$ layers[31].

The pore-free composite was obtained after co-electroplating, suitable to prepare atom probe specimens, as shown in Figure 2b. APT specimens were prepared using a FEI Quanta 3D dual-beam scanning-electron microscope/focused-ion beam (SEM/FIB) and following the in situ lift-out protocol outlined in ref.[32] and detailed in Figures 2c–f. The atom probe data were acquired on a CAMECA LEAP 5000HR at a laser pulse frequency of 125 kHz, a base temperature of 60 K, a laser pulse energy of 80 pJ, and a detection rate of 0.5%.

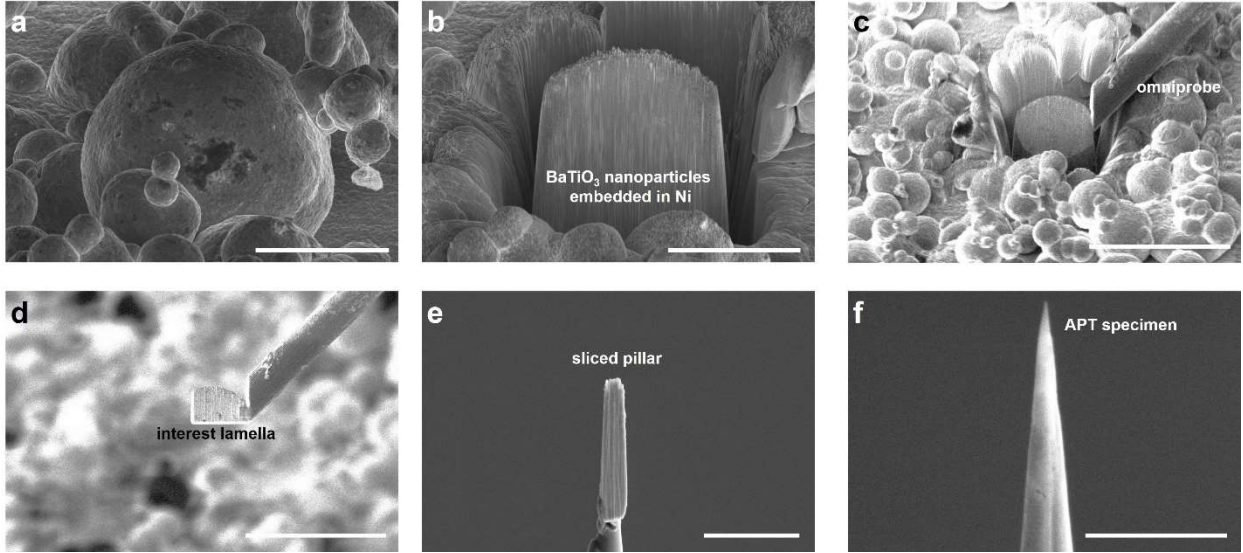

**Figure 2.** APT specimen preparation for the Ni-coated BaTiO$_3$ powder sample: (a) 52° tilted SEM image of a selected protrusion region. (b) Trenches were milled 15 μm in depths on the front/back and the left-side of the region of interest with a width of 1 μm. (c) An L-shape horizontal cut was made at the bottom and side of the sample in 52° relative to the ion-beam column, after welding on an omni-probe. (d) The sectioned lamella was lifted-out. (e) Subsequently, the sample was welded with a FIB-Pt deposition on a commercial Si micro-post on only one side where a free cut was done. (f) Thereafter, a free-cut sample was annular milled from the top in decreasing diameters until the apex radius was below 100 nm and the capped Ni layer was removed. The white scale bars are (a,b) 20, (c,d) 50, (e) 10, (f) 1 μm.

Figure 3a shows the 3D atom map of the BaTiO$_3$ particles embedded within the electroplated Ni matrix. The particle size was similar to the TEM observation. Figure 3b displays the elemental distribution for an individual BaTiO$_3$ particle. Ca and Sr are detected as well as unexpected elements such as Na and Sc. Eiser and Beck had previously reported contamination by Sr and Ca in a similar material[33]. 1$^{st}$ order nearest-neighbor analysis of the major impurities, in Figure 3c, show that they are randomly distributed, except for Na. A cylindrical region-of-interest was positioned normal to the particle's interface with the metallic matrix. The corresponding composition profile plotted in Figure 3d indicates neither chemical partitioning nor segregation of the major elements, but Na shows an enrichment on the particle's surface.

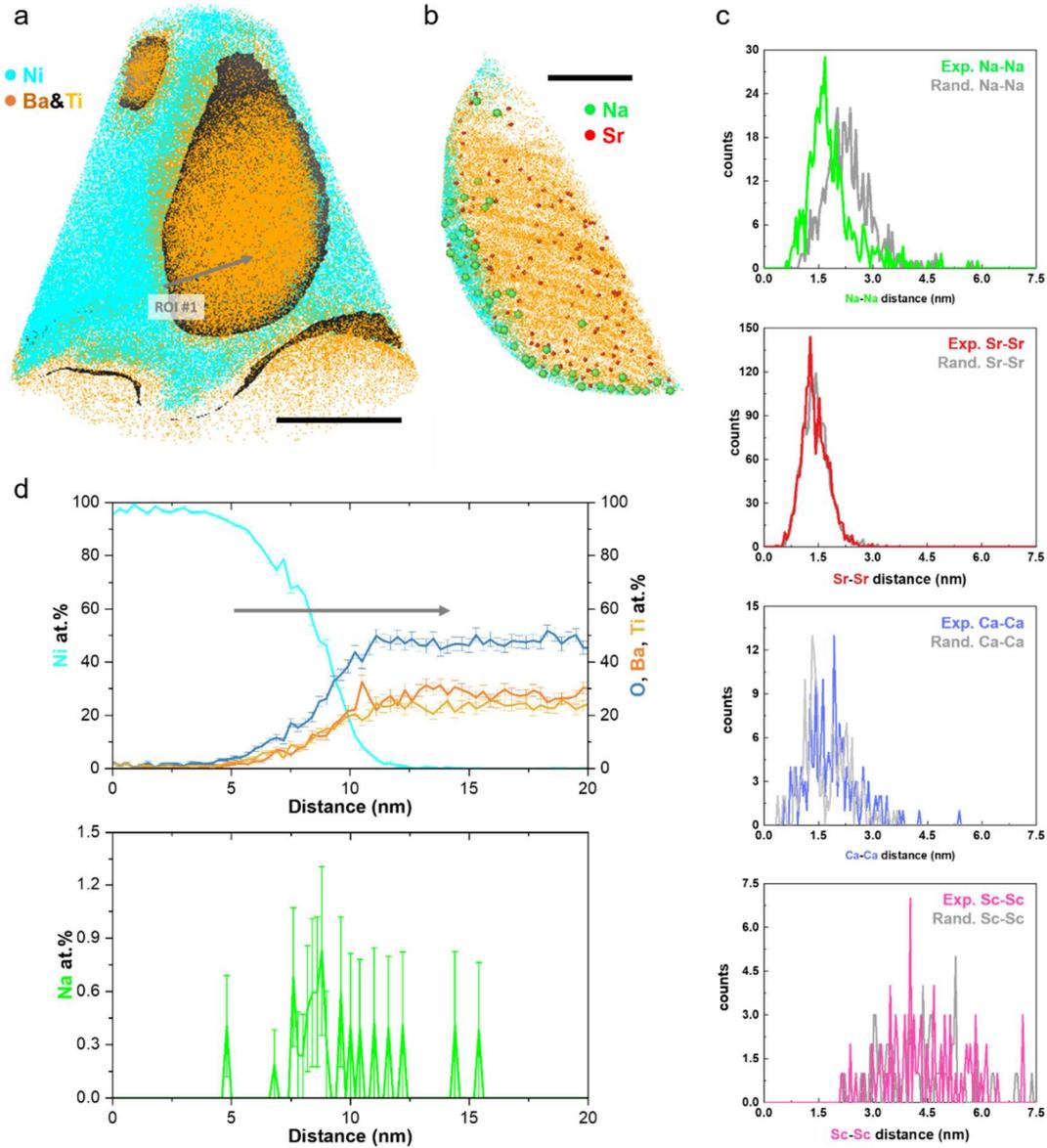

**Figure 3.** (a) 3D atom map of BaTiO$_3$ particles embedded in Ni. (b) Distribution of constituent elements (Ba, Ti, Na, Sr) extracted from the atom map. (a,b) scale bars are 20 nm. (c) Nearest-neighbor displacement analyses of Na, Sr, Ca, and Sc elements. (d) 1D compositional analysis along the particle surface. A gray arrow marks the interest profile.

To rationalize why the analysis of individual particles of $BaTiO_3$ embedded in Ni is successful, we performed GGA-level spin-polarized density-functional theory calculations with the Vienna ab initio simulation package (VASP) using a plane wave basis set with cut-off energy of 400 eV. The hybrid functional HSE06 was used to describe the electronic structure accurately[34] (see the SI for computational details).

Figure 4a shows the charge density difference in the bare $BaTiO_3$ between with and without an applied electric field. The 2D mapping color codes the electron transfer processes, with electron-rich (in red) and electron-poor (in blue), with the corresponding atomic positions displayed on the side. The linear profile of the charge density change quantifies a very subtle change under an electric field of 0.05 eV/Å, with no localization of charges and a polarized material. The spontaneous polarization of $BaTiO_3$ along the direction of the applied electric field leads to lattice distortions, with a Ti off-centered displacement along the c-axis from the tetragonal structure with a rate of 149 pm $V^{-1}$[35]. With the conductive Ni layer on the $BaTiO_3$, shown in Figure 4b, charges flow from the $BaTiO_3$ layers into the Ni, reverting the polarizations and shielding the external field. These calculations indicate that the conductive coating prevent field penetration as expected.

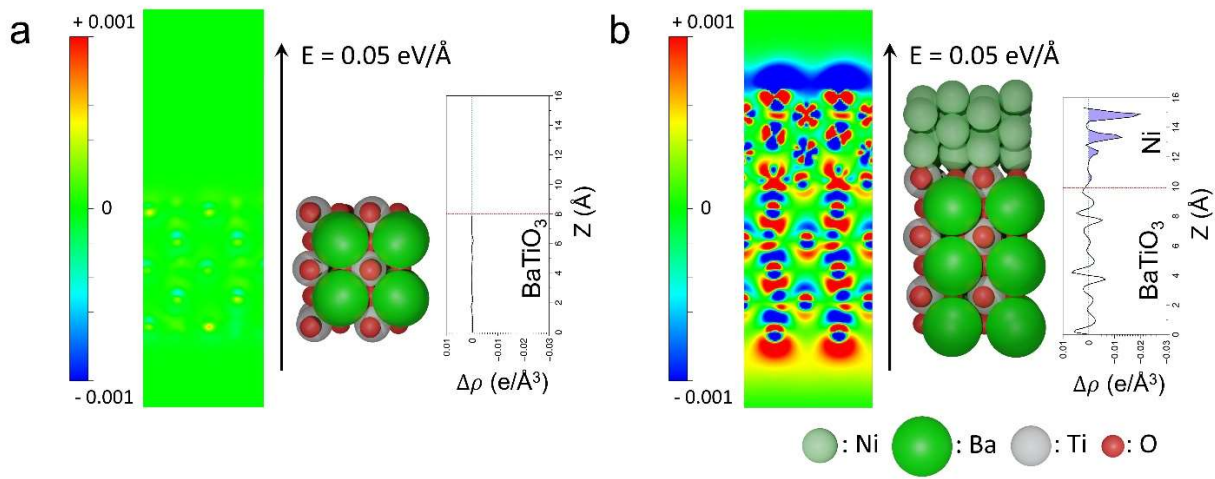

**Figure 4.** Charge distributions under an applied electric field along the z-direction for (a) bare BaTiO$_3$ and (b) Ni-coated BaTiO$_3$. Each figure includes the charge-density change distribution before/after the applied field mapped in 2D with color mapping (left), illustrated structure model in identical atomic positions (center), and an electronic density changes along the z-direction (right).

Typical field evaporation occurs at an electrostatic potential of a few kV, and under fields in the range of tens of V·nm$^{-1}$ that cause strong electrostatic Maxwell stresses, making APT studies of brittle and non-conductive materials intrinsically challenging[36–39]. Additionally, most perovskite materials exhibit intrinsic piezoelectricity that is an electromechanical interaction between mechanical and electrical states; a polarization from external electric field generates an internal mechanical load to the material. The combination of these factors, at the low temperature at which APT analyses are typically performed (25–80K) facilitate early, brittle fracture of the specimen[40,41]. Often heavily doped perovskites were successfully measured with APT, which could be attributed to the modified electronic structure[42], yet low yields had been reported[43]. The metallic coating suppressed the expression of the intrinsic piezoelectricity of BaTiO$_3$, which we believe explains the successful application of APT. Besides Ni, Co can be selected as a charge-shielding material for possible other candidate, as it is a monoisotopic element and has a higher evaporation field than Ni[44]. The use of experimental conditions during APT analysis that would minimize the electrostatic field could facilitate successful data acquisition, *i.e.* a higher temperature raises from the pulsed laser illumination that could arise from the use of a shorter wavelength for instance.

A locally exposed surface of BaTiO3 may experience piezoelectricity, however only from the one exposed surface, and not from all around a complete specimen. In addition, here, the surrounding metal matrix can deform more easily under the influence of the electrostatic pressure arising from the electric field. When there were more than two nanoparticles simultaneously imaged on the detector, the specimen immediately fractured. This suggests that the ratio of the exposed surface of BaTiO3 nanoparticles to Ni matrix is an essential factor for successful measurement.

The atomic-scale elemental mapping of $BaTiO_3$ helps identifying impurities on the particles' surface that could be problematic in achieving reliable high-yield manufacturing. During sintering, grain boundaries form along interfaces between adjacent particles, removing pores by surface diffusion at high temperature[45]. Impurities at the particle's surface can be trapped and decorate the grain boundaries, thereby modifying their properties[46]. In recent studies on the grain boundary engineering, the residual impurities on the colloidal nanocrystals surface segregated at grain boundaries during coarsening eventually change the nanomaterial's charge transport properties[47,48]. As Na acts as a dopant in $BaTiO_3$[28,49] tuning the material's dielectric constant, ferromagnetic, and optical properties, the surface contamination on the pre-processed particles can be another factor that affect the intrinsic property of as-sintered $BaTiO_3$, calling for great caution.

**Acknowledgments**

S.-H.K. and B.G. acknowledge financial support from the DFG through DIP Project No. 450800666. X.Z. is grateful for financial support from the Alexander von Humboldt Foundation. Theoretical calculation was supported by the National Research Foundation of Korea (NRF) funded by the Ministry of Education (NRF-2021R1A6A1A03043682). The calculations were supported by the Welch Foundation (F-1841) and the Texas Advanced Computing Center.